\documentclass[aoas,seceqn,mtplusscr,dvips]{arximspdf}
\usepackage{mathrsfs}
\usepackage[psamsfonts]{euscript}
\usepackage{graphicx}

\doi{10.1214/09-AOAS312}
\volume{3}
\issue{4}
\pubyear{2009}
\firstpage{1236}
\lastpage{1265}

\makeatletter

 \newtheorem{theorem}{Theorem}
\newtheorem{lemma}{Lemma}
\newtheorem{corollary}{Corollary}
 \newproclaim{remark}{Remark}
 \newproclaim{definition}{Definition}
 \newproclaim{example}{Example}

\makeatother

\begin{document}
\begin{frontmatter}

\title{Brownian distance covariance}
\runtitle{Brownian covariance}
\relateddoi{}{Discussed in \doi{10.1214/09-AOAS312A},
\doi{10.1214/09-AOAS312B},
\doi{10.1214/09-AOAS312C},
\doi{10.1214/09-AOAS312D},
\doi{10.1214/09-AOAS312E},
\doi{10.1214/09-AOAS312F}
and
\doi{10.1214/09-AOAS312G};
rejoinder at
\doi{10.1214/09-AOAS312REJ}.}

\begin{aug}
\author[A]{\fnms{G\'{a}bor J.} \snm{Sz\'{e}kely}\ead[label=e2]{gszekely@nsf.gov}\thanksref{t1}}
\and
\author[B]{\fnms{Maria L.} \snm{Rizzo}\ead[label=e1]{mrizzo@bgsu.edu}\corref{}}
\thankstext{t1}{Research supported in part by the NSF.}
\runauthor{G. J. Sz\'{e}kely and M. L. Rizzo}
\affiliation{Bowling Green State University, Hungarian Academy of Sciences and Bowling~Green State University}
\address[A]{Department of Mathematics and Statistics\\
Bowling Green State University\\
Bowling Green, Ohio 43403\\USA\\
and\\
R\'{e}nyi Institute of Mathematics\\
Hungarian Academy of Sciences\\
Budapest\\
Hungary\\
\printead{e2}} 
\address[B]{Department of Mathematics and Statistics\\
Bowling Green State University\\
Bowling Green, Ohio 43403\\USA\\
\printead{e1}}
\end{aug}

\received{\smonth{6} \syear{2009}}
\revised{\smonth{10} \syear{2009}}

%
\begin{abstract}
Distance correlation is a new class of multivariate dependence coefficients
applicable to random vectors of arbitrary and not necessarily equal dimension.
Distance covariance and distance correlation are analogous to
product-moment covariance and correlation,
but generalize and extend these classical bivariate measures of
dependence. Distance correlation characterizes
independence: it is zero if and only if the random vectors are
\mbox{independent}. The notion of covariance with respect
to a stochastic process is introduced, and it is shown that population
distance covariance coincides with the
covariance with respect to Brownian motion; thus, both can be called
Brownian distance covariance. In the bivariate case,
Brownian covariance is the natural extension of product-moment
covariance, as we obtain Pearson product-moment covariance
by replacing the Brownian motion in the definition with identity. The
corresponding statistic has an elegantly simple computing
formula. Advantages of applying Brownian covariance and correlation vs
the classical Pearson covariance and correlation are
discussed and illustrated.
\end{abstract}

%
\begin{keyword}
\kwd{Distance correlation}
\kwd{dcor}
\kwd{Brownian covariance}
\kwd{independence}
\kwd{multivariate}.
\end{keyword}

\end{frontmatter}

\section{Introduction}

The importance of independence arises in diverse applications, for
inference and whenever it is essential to measure complicated
dependence structures in bivariate or multivariate data. This paper
focuses on a new dependence coefficient that measures all types of
dependence between random vectors $X$ and $Y$ in arbitrary dimension.
\emph{Distance correlation} and \emph{distance covariance} (Sz\'
{e}kely, Rizzo, and Bakirov \cite{srb07}), and \emph{Brownian
covariance}, introduced in this paper, provide a new approach to the
problem of measuring dependence and testing the joint independence of
random vectors in arbitrary dimension.
The corresponding statistics have simple computing formulae, apply to
sample sizes $n \geq2$ (not constrained by dimension), and do not
require matrix inversion or estimation of parameters. For example, the
distance covariance (dCov) statistic, derived in the next section, is
the square root of
\[
\mathcal V_n^2 = {\frac{1}{n^2} \sum_{k, l=1}^n A_{kl}B_{kl}} ,
\]
where $A_{kl}$ and $B_{kl}$ are simple linear functions of the pairwise
distances between sample elements. It will be shown that the
definitions of the new dependence coefficients have theoretical
foundations based on characteristic functions and on the new concept of
covariance with respect to Brownian motion. Our independence test
statistics are consistent against all types of dependent alternatives
with finite second moments.

Classical Pearson product-moment correlation ($\rho$) and covariance
measure linear dependence between two random variables, and in the
bivariate normal case $\rho=0$ is equivalent to independence. In the
multivariate normal case, a diagonal covariance matrix $\Sigma$
implies independence, but is not a sufficient condition for
independence in the general case. Nonlinear or nonmonotone dependence
may exist. Thus, $\rho$ or $\Sigma$ do not characterize independence
in general.

Although it does not characterize independence, classical correlation
is widely applied in time series, clinical trials, longitudinal
studies, modeling financial data, meta-analysis, model selection in
parametric and nonparametric models, classification and pattern
recognition, etc. Ratios and other methods of combining and applying
correlation coefficients have also been proposed. An important example
is maximal correlation, characterized by R\'{e}nyi \cite{renyi59}.

For multivariate inference, methods based on likelihood ratio tests
(LRT) such as Wilks' Lambda \cite{wilks} or Puri-Sen \cite{ps} are
not applicable if dimension exceeds sample size, or when distributional
assumptions do not hold. Although methods based on ranks can be applied
in some problems, many classical methods are effective only for testing
linear or monotone types of dependence.

There is much literature on testing or measuring independence. See, for
example, Blomqvist \cite{blom50}, Blum, Kiefer, and Rosenblatt \cite
{bkr61}, or methods outlined in Hollander and Wolfe \cite{hw99} and
Anderson \cite{anderson03}. Multivariate nonparametric approaches to
this problem can be found in Taskinen, Oja, and Randles \cite{tor05},
and the references therein.

Our proposed distance correlation represents an entirely new approach.
For all distributions with finite first moments, \emph{distance
correlation} $\mathcal R$ generalizes the idea of \emph{correlation}
in at least two fundamental
ways:
\begin{longlist}[(ii)]
\item[(i)] $\mathcal R(X,Y)$ is defined for $X$ and $Y$ in arbitrary dimension.
\item[(ii)] $\mathcal R(X,Y)=0$ characterizes independence of $X$ and $Y$.
\end{longlist}
The coefficient $\mathcal R(X,Y)$ is a standardized version of distance
covariance $\mathcal V(X,Y)$, defined in the next section. Distance
correlation satisfies $0 \le\mathcal R \le1$, and $\mathcal R =0$
only if $X$ and $Y$ are independent. In the \emph{bivariate normal}
case, $\mathcal R$ is a deterministic function of $\rho$, and
$\mathcal{R}(X,Y) \leq|\rho(X,Y)|$ with equality when $\rho=\pm1$.

Thus, distance covariance and distance correlation provide a natural
extension of Pearson product-moment covariance $\sigma_{X,Y}$ and
correlation $\rho$, and new methodology for measuring dependence in
all types of applications.

The notion of covariance of random vectors $(X,Y)$ with respect to a
stochastic process $U$ is introduced in this paper. This new notion
$\operatorname{Cov}_U(X,Y)$ contains as distinct special cases distance covariance
$\mathcal V^2(X,Y)$ and, for bivariate $(X,Y)$, $\sigma_{X,Y}^2$. The
title of
this paper refers to $\operatorname{Cov}_W(X,Y)$, where $W$ is a Wiener process.

\emph{Brownian covariance} $\mathcal W=\mathcal W(X,Y)$ is based on
 Brownian motion or Wiener process for random variables $X\in
\mathbb R^p$ and $Y\in\mathbb R^q$ with finite second moments. An
important property of Brownian covariance is that $\mathcal W(X,Y)=0$
if and only if $X$ and $Y$ are independent.

A surprising result develops: the Brownian covariance is equal to the
distance covariance. This equivalence is not only surprising, it also
shows that distance covariance is a natural counterpart of
product-moment covariance. For bivariate $(X,Y)$, by considering the
simplest nonrandom function, identity ($\mathit{id}$), we obtain $\operatorname{Cov}
_{\mathit{id}}(X,Y)=\sigma_{X,Y}^2$. Then by considering the most fundamental random
processes, Brownian motion $W$, we arrive at $\operatorname{Cov}_W(X,Y)=\mathcal V^2(X,Y)$.
Brownian correlation is a standardized Brownian covariance, such that
if Brownian motion is replaced with the identity function, we obtain
the absolute value of Pearson's correlation $\rho$.

A further advantage of extending Pearson correlation with distance
correlation is that while uncorrelatedness ($\rho=0$) can sometimes
replace independence, for example, in proving some classical laws of
large numbers, uncorrelatedness is too weak to imply a central limit
theorem, even for strongly stationary summands (see Bradley
\cite{bradley81,bradley88,bradley07}). On the other hand, a central limit
theorem for strongly stationary sequences of summands follows from
$\mathcal R = 0$ type conditions (Sz\'{e}kely and Bakirov \cite{TR08}).

Distance correlation and distance covariance are presented in Section
\ref{s:dcor}. Brownian covariance is introduced in Section \ref
{sec3}. Extensions and applications are discussed in Sections \ref
{sec4} and \ref{sec5}.

\section{Distance covariance and distance correlation}\label{s:dcor}

Let $X $ in $ \mathbb{R}^p$ and $Y $ in $ \mathbb{R}^q$ be random
vectors, where $p$ and $q$ are positive integers. The lower case $f_{X}$
and $f_{Y}$ will be used to denote the characteristic functions of $X$
and $Y$, respectively, and their joint characteristic function is
denoted $f_{X,Y}$. In terms of characteristic functions, $X$ and $Y$ are
independent if and only if $f_{X,Y}=f_{X}f_{Y}$. Thus, a natural approach
to measuring the dependence between $X$ and $Y$ is to find a suitable
norm to measure the distance between $f_{X,Y}$ and $f_{X}f_{Y}$.

Distance covariance $\mathcal V$ is a measure of the distance between
$f_{X,Y}$ and the product $f_{X}f_{Y}$. A norm $\| \cdot\|$ and a distance
$\| f_{X,Y}-f_{X}f_{Y}\|$ are defined in Section \ref{defs1}. Then an
empirical version of $\mathcal V$ is developed and applied to test the
hypothesis of independence
\begin{eqnarray*} 
H_0\dvtx f_{X,Y}=f_{X}f_{Y}\quad \mbox{vs}\quad H_1\dvtx f_{X,Y}\neq f_{X}f_{Y}.
\end{eqnarray*}

In Sz\'{e}kely et al. \cite{srb07} an omnibus test of independence
based on the sample distance covariance $\mathcal V$ is introduced that
is easily implemented in arbitrary dimension without requiring
distributional assumptions. In Monte Carlo studies, the distance
covariance test exhibited superior power relative to parametric or
rank-based likelihood ratio tests against nonmonotone types of
dependence. It was also demonstrated that the tests were quite
competitive with the parametric likelihood ratio test when applied to
multivariate normal data. The practical message is that distance
covariance tests are powerful tests for all types of dependence.

\subsection{Motivation}

\subsubsection*{Notation} The scalar product of vectors $t$ and $s$ is
denoted by
$\langle t,s \rangle$. For complex-valued functions $f(\cdot)$, the
complex conjugate of $f$ is denoted by $\overline{f}$ and
$|f|^2=f \overline{f}.$ The Euclidean norm of $x$ in $ \mathbb
R^p$ is $|x|_p$. A primed variable $X'$ is an independent copy of $X$;
that is, $X$ and $X'$ are independent and identically distributed (i.i.d.).

For complex functions $\gamma$ defined on $\mathbb{R}^p \times
\mathbb{R}^q$, the $\Vert \cdot\Vert_w$-norm in the weighted $L_2$
space of functions on $\mathbb{R}^{p+q}$ is defined by
%
\begin{equation}\label{wnorm}
{ \|\gamma(t,s) \|}_w^2= \int_{\mathbb{R}^{p + q}}|\gamma(t,s)|^2
w (t, s) \,dt\, ds,
\end{equation}
where $w(t,s)$ is an arbitrary positive weight function for which
the integral above exists.

With a suitable choice of weight function $w(t,s)$, discussed below, we
shall define a measure of dependence
\begin{eqnarray}\label{e:Aw}
\mathcal{V}^2(X,Y;w)&=& \| f_{X,Y}(t,s)-f_{X}(t)f_{Y}(s) \|_w^2\nonumber
\\[-8pt]\\[-8pt]
&=&
\int_{\mathbb{R}^{p + q}}|f_{X,Y}(t,s) - f_{X}(t)f_{Y}(s)|^2 w (t, s)
\,dt\, ds,\nonumber
\end{eqnarray}
which is analogous to classical covariance, but with the important
property that $\mathcal{V}^2(X,Y;w)=0$ if and only if $X$ and $Y$ are
independent. In what follows, $w$ is chosen such that we can also define
\begin{eqnarray*}
\mathcal{V}^2(X;w) &=& \mathcal{V}^2(X,X;w)= \| f_{X,X}(t,s)-f_{X}(t)f_{X}
(s) \|_w^2
\\
&=&
\int_{\mathbb{R}^{2p}}|f_{X,X}(t,s) - f_{X}(t)f_{X}(s)|^2 w (t, s)
\,dt\, ds,
\end{eqnarray*}
and similarly define $\mathcal{V}^2(Y;w)$.
Then a standardized version of $\mathcal{V}(X,Y;w)$ is
\[
\mathcal R_w = \frac{\mathcal V(X,Y;w)}{\sqrt{\mathcal V (X;w)
\mathcal V (Y;w)}},
\]
a type of unsigned correlation.

In the definition of the norm (\ref{wnorm}) there are more than one
potentially interesting and applicable choices of weight function $w$,
but not every $w$ leads to a dependence measure that has desirable
statistical properties. Let us now discuss the motivation for our
particular choice of weight function leading to distance covariance.

At least two conditions should be satisfied by the standardized
coefficient $\mathcal R_w$:
\begin{longlist}[(ii)]
\item[(i)] $\mathcal R_w \geq0$ and $\mathcal R_w=0$ only if independence holds.
\item[(ii)] $\mathcal R_w$ is scale invariant, that is, invariant with
respect to transformations $(X,Y) \mapsto(\epsilon X, \epsilon Y)$,
for $\epsilon> 0$.
\end{longlist}
However, if we consider \emph{integrable} weight function $w(t,s)$,
then for $X$ and $Y$ with finite variance
\begin{eqnarray*}
\lim_{\epsilon\to0} \frac{\mathcal{V}^2 (\epsilon X, \epsilon
Y;w)}{\sqrt{
\mathcal{V}^2(\epsilon X;w)\mathcal{V}^2(\epsilon Y;w)}} =
\rho^2(X,Y).
\end{eqnarray*}
The above limit is obtained by considering the Taylor expansions of the
underlying characteristic functions. Thus, if the weight function is
integrable, $\mathcal R_w$ can be arbitrarily close to zero even if $X$
and $Y$ are dependent. By using a suitable \emph{nonintegrable} weight
function, we can obtain an $\mathcal R_w$ that satisfies both
properties (i) and (ii) above.

Considering the operations on characteristic functions involved in
evaluating the integrand in (\ref{e:Aw}), a promising solution to the
choice of weight function $w$ is suggested by the following lemma.

\begin{lemma} \label{lemmaC}
If $0<\alpha<2$, then for all ${x}$ in $ \mathbb R^d$
\[
\int_{\mathbb R^d}\frac{1- \cos\langle{t},{x} \rangle}{|{
t}|_d^{d+\alpha}} \, d{t}=C(d, \alpha)|{x}|_d^{\alpha} ,\quad
\]
where
\begin{eqnarray*}\label{e:C}
C(d,\alpha)&=\dfrac{2{\pi}^{{d/2}} \Gamma( 1-{{\alpha}/2})}
{\alpha2^{\alpha}\Gamma( {(d+\alpha)/2})},
\end{eqnarray*}
and $\Gamma(\cdot)$ is the complete gamma function. The integrals at
$0$ and $\infty$ are meant in the principal value sense: $
\lim_{\varepsilon\to0}\int_{{\mathbb R}^d
\setminus
\{\varepsilon B +{\varepsilon}^{-1} {B^c}\}},
$
where $B$ is the unit ball (centered at 0) in ${\mathbb R}^d$ and
${B^c}$ is the complement of $B$.
\end{lemma}

A proof of Lemma \ref{lemmaC} is given in Sz\'{e}kely and Rizzo \cite{sr05b}.
Lemma \ref{lemmaC} suggests the weight functions
%
\begin{equation}\label{wpq}
w(t,s;\alpha) = (C(p,\alpha)C(q,\alpha) |t|_p^{p+\alpha
}|s|_q^{q+\alpha} )^{-1}, \qquad 0<\alpha<2.
\end{equation}
The weight functions (\ref{wpq}) result in coefficients $\mathcal R_w$
that satisfy the scale invariance property (ii) above.

In the simplest case corresponding to $\alpha=1$ and Euclidean norm $|x|$,
%
\begin{equation}\label{w}
w(t,s) = (c_p c_q |t|_p^{1+p}|s|_q^{1+q} )^{-1},
\end{equation}
where
%
\begin{equation}\label{e:C1}
c_d = C(d,1) = \frac{{\pi}^{{(1+d)/2}}}{ \Gamma( {(1+d)/2})}.
\end{equation}
(The constant $2c_d$ is the surface area of the unit sphere in $\mathbb
R^{d+1}$.)

\begin{remark}
Lemma \ref{lemmaC} is applied to evaluate the integrand in (\ref
{e:Aw}) for weight functions (\ref{wpq}) and (\ref{w}). For example, if
$\alpha=1$ (\ref{w}), then by Lemma \ref{lemmaC} there exist
constants $c_p$
and $c_q$ such that for $X $ in $ \mathbb{R}^p$ and $Y$ in
$\mathbb{R}^q$,
\begin{eqnarray*}
&\displaystyle \int_ {\mathbb R^p}
\frac{1-\exp\{i \langle t, X \rangle\}}
{|t|_p^{1+p} }
\,dt = c_{p} | X |_p , \qquad
\int_ {\mathbb R^q}
\frac{1-\exp\{i \langle s, Y \rangle\}}
{|s|_q^{1+q} }
\,ds = c_{q} | Y |_q ,&
\\
& \displaystyle\int_ {\mathbb R^p} \int_ {\mathbb R^q}
\frac{1-\exp\{i \langle t, X \rangle+ i \langle s, Y \rangle\}}
{|t|_p^{1+p} |s|_q^{1+q}}
\,dt \,ds = c_{p} c_{q} | X |_p | Y |_q.&
\end{eqnarray*}
\end{remark}

Distance covariance and distance correlation are a class of dependence
coefficients and statistics obtained by applying a weight function of
the type (\ref{wpq}), $0 < \alpha< 2$. This type of weight function
leads to a simple product-average form of the covariance (\ref
{e:AnXY}) analogous to Pearson covariance. Other interesting weight
functions could be considered (see, e.g., Bakirov, Rizzo and Sz\'
{e}kely \cite{brs06}), but only the weight functions (\ref{wpq}) lead
to distance covariance type statistics (\ref{e:AnXY}).

In this paper we apply weight function (\ref{w}) and the corresponding
weighted $L_2$ norm $\| \cdot\|$, omitting the index $w$, and write
the dependence measure (\ref{e:Aw}) as $\mathcal{V}^2(X,Y)$. Section
\ref{alpha} extends our results for $\alpha\in(0,2)$.

For finiteness of $\| f_{X,Y}(t,s)- f_{X}(t)f_{Y}(s)\|^2 $, it is
sufficient that $E|X|_p<\infty$ and $E|Y|_q<\infty$.

\subsection{Definitions}\label{defs1}

\begin{definition}\label{defdCov}
The \textit{distance covariance} ($\operatorname{dCov}$)
between random
vectors $X $ and $Y
$ with finite first moments is the nonnegative number
$\mathcal V(X,Y)$ defined by
\begin{eqnarray}\label{Adef}
\mathcal{V}^2(X,Y) &=&
\| f_{X,Y}(t,s)-
f_{X}(t)f_{Y}(s)\|^2\nonumber
\\[-8pt]\\[-8pt]
&=& \frac{1}{c_pc_q}
\int_{\mathbb{R}^{p+q}}
\frac{| f_{X,Y}(t,s)-
f_{X}(t)f_{Y}(s)|^2} {|t|_p^{1+p} |s|_q^{1+q}}\, dt \,ds.\nonumber
\end{eqnarray}
\end{definition}

Similarly, distance variance ($\operatorname{dVar}$) is defined as the
square root
of
\[
\mathcal V^2(X)=\mathcal{V}^2(X,X)=\| f_{X,X}(t,s)-
f_{X}(t)f_{X}(s)\|^2.
\]

By definition of the norm $\| \cdot\|$, it is clear that $\mathcal
V(X,Y) \geq0$ and $\mathcal V(X,Y)=0$ if and only if $X$ and $Y$ are
independent.

\begin{definition}
The \textit{distance correlation} ($dCor$) between random
vectors $X $ and $Y
$ with finite first moments is the nonnegative number
$\mathcal{R}(X,Y)$ defined by
%
\begin{equation}\label{e:alpha}
\mathcal{R}^2(X,Y) =
\cases{
\dfrac{\mathcal{V}^2(X,Y)}{\sqrt{\vphantom{\int}\mathcal{V}^2(X)
\mathcal{V}^2(Y)}},
& \quad$\mathcal{V}^2(X)\mathcal{V}^2(Y)>0$;
\cr
0, & \quad $\mathcal{V}^2(X)\mathcal{V}^2(Y)=0$.
}
\end{equation}
\end{definition}

Several properties of $\mathcal R$ analogous to $\rho$ are given in
Theorem \ref{Th3}. Results for the special case of bivariate normal
$(X,Y)$ are given in Theorem \ref{T:abvn}.

The distance dependence statistics are defined as follows. For a
random sample $(\mathbf{X,Y})=\{(X_k, Y_k)\dvtx k=1,\ldots,n\}$ of $n$ i.i.d.
random vectors $(X,Y)$ from the joint distribution of random vectors $X
$ in $
\mathbb{R}^p$ and $Y $ in $ \mathbb{R}^q$, compute the Euclidean
distance matrices $(a_{kl})=(|X_k-X_l|_p)$ and
$(b_{kl})=(|Y_k-Y_l|_q)$. Define
\begin{eqnarray*}
A_{kl}= a_{kl}-\bar a_{k\cdot}- \bar a_{\cdot l} + \bar a_{\cdot\cdot} ,
\qquad k,l=1,\ldots,n,
\end{eqnarray*}
where
\begin{eqnarray*}
\bar a_{k \cdot}= \frac{1}{n}
\sum_{l=1}^n a_{kl}, \qquad\bar a_{\cdot l}, = \frac{1}{n}
\sum_{k=1}^n
a_{kl}, \qquad
\bar a_{\cdot\cdot} = \frac{1}{n^2} \sum_{k, l=1}^n a_{kl}.
\end{eqnarray*}
Similarly, define
$B_{kl}= b_{kl} -\bar b_{k \cdot}- \bar b_{\cdot l} + \bar b_{\cdot\cdot}$, for
$k,l=1,\ldots,n$.

\begin{definition}\label{defVn}
The nonnegative sample distance covariance $\mathcal{V}_n(\mathbf{X},\mathbf
{Y})$ and sample distance correlation $\mathcal R_n(\mathbf X, \mathbf
Y)$ are defined by
%
\begin{equation}\label{e:AnXY}
\mathcal{V}^2_n (\mathbf{X,Y}) = {\frac{1}{n^2} \sum_{k, l=1}^n
A_{kl}B_{kl}}
\end{equation}
and
%
\begin{equation}\label{e:alphan}
\mathcal{R}^2_n(\mathbf{X,Y})=
\cases{
\dfrac{\mathcal{V}^2_n(\mathbf{X,Y})}{\sqrt{\vphantom{\overline{A}}
\mathcal{V}^2_n (\mathbf{X}) \mathcal{V}^2_n(\mathbf{Y})}} ,
& \quad $ \mathcal{V}^2_n
(\mathbf{X}) \mathcal{V}^2_n (\mathbf{Y}) >0
$; \cr
0, &\quad $\mathcal{V}^2_n
(\mathbf{X}) \mathcal{V}^2_n (\mathbf{Y}) =0$,
}
\end{equation}
respectively, where the sample distance variance is defined by
%
\begin{equation}\label{e:AnX}
\mathcal{V}^2_n (\mathbf{X}) = \mathcal{V}^2_n (\mathbf{X,X}) =
{\frac{1}{n^2} \sum_{k, l=1}^n
A_{kl}^2} .
\end{equation}
\end{definition}

The nonnegativity of $\mathcal{R}_n^2$ and $\mathcal{V}_n^2$ may not be
immediately obvious from the definitions above, but this property as
well as the motivation for the definitions of the statistics will
become clear from Theorem
\ref{TH1} below.

\subsection{Properties of distance covariance}

Several interesting properties of distance covariance are obtained.
Results in this section are summarized as follows:
\begin{longlist}[(iii)]
\item[(i)] Equivalent definition of $\mathcal V_n$ in terms of empirical
characteristic functions and norm $\| \cdot\|$.
\item[(ii)] Almost sure convergence $\mathcal V_n \to\mathcal V$ and
$\mathcal R_n^2 \to\mathcal R^2$.
\item[(iii)] Properties of $\mathcal V(X,Y)$, $\mathcal V(X)$, and $\mathcal R(X,Y)$.
\item[(iv)] Properties of $\mathcal R_n$ and $\mathcal V_n$.
\item[(v)] Weak convergence of $n\mathcal V_n^2$, the limit distribution of
$n\mathcal V_n^2$, and statistical consistency.
\item[(vi)] Results for the bivariate normal case.
\end{longlist}
Many of these results were obtained in Sz\'{e}kely et al. \cite
{srb07}. Here we give the proofs of new results and readers are
referred to \cite{srb07} for more details and proofs of our previous
results.

\subsubsection*{An equivalent definition of $\mathcal V_n$}
The coefficient $\mathcal V(X,Y)$ is defined in terms of characteristic
functions, thus, a natural approach is to define the statistic
$\mathcal V_n(\mathbf X, \mathbf Y)$ in terms of empirical
characteristic functions. The joint empirical characteristic function
of the sample,
$\{(X_1,Y_1), \ldots,(X_n,Y_n)\}$, is
\[
f_{X,Y}^n(t, s) = \frac1{n} \sum_{k = 1}^n\exp\{i \langle t, X_k
\rangle+
i \langle s, Y_k \rangle\}.
\]
The marginal empirical characteristic functions of the $X$
sample and $Y$ sample are
\[
f_{X}^{n} (t) = \frac1{n} \sum_ {k = 1} ^ n\exp\{i \langle t, X _ k
\rangle\},
\qquad
f_{Y}^{n} (s) = \frac1{n} \sum_ {k = 1} ^ n\exp\{i \langle s, Y _ k
\rangle
\},
\]
respectively. Then an empirical version of distance covariance could
have been
defined as $ \| f_{X,Y}^n(t,s) - f_{X}^n(t) f_{Y}^n(s) \|,
$
where the norm $\| \cdot\|$ is defined by the integral as above in
(\ref{wnorm}). Theorem \ref{TH1} establishes that this definition is
equivalent to Definition \ref{defVn}.
\begin{theorem}\label{TH1}
If $(\mathbf{X,Y})$ is a sample from the joint
distribution of $(X,Y)$, then
\begin{eqnarray*}\mathcal{V}^2_n(\mathbf{X,Y}) = \| f_{X,Y}^n(t, s)
-{f_{X}^n}(t) {f_{Y}^n}(s) \|^2.
\end{eqnarray*}
\end{theorem}

The proof applies Lemma \ref{lemmaC} to evaluate the integral $ \|
f_{X,Y}^n(t, s) -{f_{X}^n}(t)
{f_{Y}^n}(s) \|^2$ with
$w(t,s)=\{c_p c_q|t|_p^{1+p}|s|_q^{1+q}\}^{-1}$. An intermediate
result is
%
\begin{equation}\label{Sformula}
\| f_{X,Y}^n(t, s) -{f_{X}^n}(t) {f_{Y}^n}(s) \|^2 =
T_1 + T_2 - 2T_3,
\end{equation}
where
\begin{eqnarray*}
T_1&=&\frac1{n^2} \sum_{k,l =1}^n |X_k-X_l|_p |Y_k-Y_l|_q,
\\
T_2&=&\frac1{n^2} \sum_{k,l =1}^n |X _ k-X_l|_p
\frac1{n^2} \sum_{k,l=1}^n |Y _ k-Y_l|_q,
\\
T_3&=&\frac1{n^3} \sum_{k =1}^n \sum_{l,m =1}^n|X _ k-X_l|_p |Y
_k-Y_m|_q.
\end{eqnarray*}
Then the algebraic identity
$
T_1+T_2-2T_3 = \mathcal{V}^2_n(\mathbf{X,Y}),
$
where $\mathcal{V}^2_n(\mathbf{X,Y})$ is given by Definition \ref
{defVn}, is established to complete the proof.

As a corollary to Theorem \ref{TH1}, we have $\mathcal V_n^2(\mathbf X, \mathbf
Y) \geq0$. It is also easy to see that the statistic $\mathcal
V_n(\mathbf{X})=0$ if and only if every sample observation is
identical. If $\mathcal V_n(\mathbf{X})=0$, then $A_{kl}=0$ for
$k,l=1,\dots,n$. Thus, $0=A_{kk}=-\overline a_{k\cdot} - \overline
a_{\cdot k} + \overline
a_{\cdot\cdot}$ implies that $\overline a_{k\cdot} = \overline
a_{\cdot k} = \overline a_{\cdot\cdot}/2$, and
\[
0=A_{kl}=a_{kl}-\overline a_{k\cdot} - \overline a_{\cdot l} +
\overline a_{\cdot\cdot}=a_{kl}=|X_k-X_l|_p,
\]
so $X_1=\cdots=X_n$.

\begin{remark}The simplicity of formula (\ref{e:AnXY}) for $\mathcal
V_n$ in Definition \ref{defVn} has practical advantages. Although the
identity (\ref{Sformula}) in Theorem \ref{TH1} provides an alternate
computing formula for $\mathcal V_n$, the original formula in
Definition \ref{defVn} is simpler and requires less computing time
(1$/$3 less time per statistic on our current machine, for sample size 100).
Reusable computations and other efficiencies possible using the simpler
formula (\ref{e:AnXY}) execute our permutation tests in 94\% to 98\%
less time, which depends on the number of replicates. It is
straightforward to apply resampling procedures without the need to
recompute the distance matrices. See Example \ref{exJack}, where a
jackknife procedure is illustrated.
\end{remark}

\begin{theorem}\label{Thm2}
If $E|X|_p<\infty$ and $E|Y|_q<\infty$, then almost surely
\begin{eqnarray*}
\lim_{n \to\infty} \mathcal V_n(\textbf{X},\textbf{Y}) =
\mathcal V(X,Y).
\end{eqnarray*}
\end{theorem}

\begin{corollary}\label{c1}
If $E(|X|_p + |Y|_q) < \infty$, then almost surely
\[
\lim_{n \to\infty} \mathcal{R}^2_n (\mathbf{X,Y}) =
\mathcal{R}^2 (X,Y).
\]
\end{corollary}

\begin{theorem}\label{Th3}
For random vectors $X \in\mathbb R^p$ and $Y \in\mathbb R^q$ such that
$E(|X|_p+|Y|_q)<\infty$, the following properties hold:
\begin{longlist}[(iii)]
\item[(i)] $0 \leq\mathcal R(X,Y) \leq1$, and $\mathcal R=0$ if and only
if $X$ and $Y$ are independent.
\item[(ii)]  $\mathcal{V}(a_1+b_1C_1X, a_2+b_2C_2Y) = \sqrt
{|\vphantom{b^b}b_1b_2|} \mathcal V(X,Y)$,
for all constant vectors $a_1 \in\mathbb R^p$, $a_2 \in\mathbb R^q$, scalars
$b_1$, $b_2$ and orthonormal matrices $C_1$, $C_2$ in $\mathbb R^p$ and
$\mathbb R^q$, respectively.
\item[(iii)]  If the random vector $(X_1,Y_1)$ is independent of the
random vector $(X_2,Y_2)$, then
\[
\mathcal{V}(X_1+X_2, Y_1+Y_2)\le\mathcal{V}(X_1,Y_1)+\mathcal
{V}(X_2,Y_2).
\]
Equality holds if and only if $X_1$ and $Y_1$ are both constants, or
$X_2$ and $Y_2$ are both constants, or
$X_1,X_2,Y_1,Y_2$ are mutually independent.
\item[(iv)] $ \mathcal V(X)=0$ implies that $X=E[X]$, almost surely.
\item[(v)] $ \mathcal V (a+bCX)=|b| \mathcal V(X)$, for all constant vectors
$a $ in $ \mathbb R^p$, scalars $b$,
and $p \times p$ orthonormal matrices $C$.
\item[(vi)]  If $X$ and $Y$ are independent, then $ \mathcal V(X+Y)\leq
\mathcal V(X)+ \mathcal V(Y)$.
Equality holds if and only if one of the random vectors $X$ or $Y$ is constant.
\end{longlist}
\end{theorem}

Proofs of statements (iii) and (vi) are given in the \hyperref[app]{Appendix}.

\begin{theorem}\label{Th4}
\begin{longlist}[(iii)]
\item[(i)] $\mathcal V_n(\mathbf X, \mathbf Y) \geq0$.
\item[(ii)] $\mathcal V_n(\mathbf X)=0$ if and only if every sample
observation is identical.
\item[(iii)]
$0 \leq\mathcal{R}_n(\mathbf X, \mathbf Y) \leq1$.
\item[(iv)]  $\mathcal{R}_n(\mathbf{X},\mathbf Y) = 1$
implies that the
dimensions of the linear subspaces spanned by $\mathbf X$ and $\mathbf
Y$ respectively are almost surely equal, and if we assume that these
subspaces are equal, then in this subspace
\[
\mathbf{Y} = a + b\mathbf{X}C
\]
for some vector $a$, nonzero real number $b,$ and orthogonal matrix $C$.
\end{longlist}
\end{theorem}

Theorem \ref{Th3} and the results below for the dCov test can be
applied in a wide range of problems in statistical modeling and
inference, including nonparametric models, models with multivariate
response, or when dimension exceeds sample size. Some applications are
discussed in Section \ref{sec5}.

\subsubsection*{Asymptotic properties of $n\mathcal V_n^2$}\label{sectionA}
A multivariate test of independence is determined by $n\mathcal V_n^2$
or $n\mathcal V_n^2/T_2$, where $T_2=\bar a_{\cdot\cdot}\bar b_{\cdot\cdot}$ is as
defined in Theorem \ref{TH1}. If we apply the latter version, it
normalizes the statistic so that asymptotically it has expected value
1. Then if $E(|X|_p+|Y|_q)<\infty$, under independence, $n\mathcal
V_n^2/T_2$ converges
in distribution to a quadratic form
%
\begin{equation}\label{e:Q1}
Q \stackrel{\EuScript D} = \sum_{j=1}^\infty\lambda_j Z_j^2,
\end{equation}
where $Z_j$ are independent standard normal random variables,
$\{\lambda_j \}$ are nonnegative constants that depend on the
distribution of $(X,Y)$, and $E[Q]=1$. A test of independence that
rejects independence for large $n \mathcal V^2_n/T_2$ (or $n \mathcal
V_n^2$) is
statistically consistent against all alternatives with finite first
moments.

In the next theorem we need only assume finiteness of first moments for
weak convergence of $n\mathcal V_n^2$ under the independence hypothesis.

\begin{theorem}[(Weak convergence)] \label{THweak}If $X$ and $Y$ are
independent and
$E(|X|_p+|Y|_q) < \infty$, then
\[
n \mathcal V_n^2 \mathop{\longrightarrow}_{n\to\infty}^D \|
\zeta(t,s)\|^2,
\]
where $\zeta(\cdot)$ is a complex-valued zero mean Gaussian random
process with covariance function
\[
R(u,u_0) = \bigl(f_{X}(t-t_0) - f_{X}(t) \overline{f_{X}(t_0)} \bigr)
\bigl( f_{Y}(s-s_0)-f_{Y}(s) \overline{f_{Y}(s_0)} \bigr),
\]
for $u=(t,s)$, $u_0=(t_0,s_0) \in\mathbb R^p \times\mathbb R^q$.
\end{theorem}

\begin{corollary} \label{CorQstar} If $E(|X|_p + |Y|_q) < \infty$, then
\begin{longlist}[(iii)]
\item[(i)] If $X$ and $Y$ are independent, then $
n\EuScript V^2_n/T_2 \mathop{\stackrel{D}{\longrightarrow}}\limits_{n\to\infty} Q
$
where $Q$ is a nonnegative quadratic form of centered Gaussian
random variables (\ref{e:Q1}) and $E[Q]=1$.
\item[(ii)] If $X$ and $Y$ are independent, then $
n\EuScript V^2_n \mathop{\stackrel{D}{\longrightarrow}}\limits_{n\to\infty} Q_1
$
where $Q_1$ is a nonnegative quadratic form of centered Gaussian
random variables and $E[Q_1]=E|X-X'| E|Y-Y'|$.
\item[(iii)] If $X$ and $Y$ are
dependent, then $
n\mathcal V^2_n/T_2 \mathop{\stackrel{P}{\longrightarrow}}\limits_{n\to\infty}
\infty$ and $
n\mathcal V^2_n \mathop{\stackrel{P}{\longrightarrow}}\limits_{n\to\infty}
\infty.
$
\end{longlist}
\end{corollary}

Corollary \ref{CorQstar}(i), (ii) guarantees that the
dCov test statistic has a proper limit distribution under the
hypothesis of independence for all $X$ and $Y$ with finite first
moments, while Corollary \ref{CorQstar}(iii) shows that under
any dependent alternative, the dCov test statistic tends to infinity
(stochastically). Thus, the dCov test of independence is statistically
consistent against all types of dependence.

The dCov test is easy to implement as a permutation test, which is the
method that we applied in our examples and power comparisons. For the
permutation test implementation one can apply test statistic $n
\mathcal V_n^2$. Large values of $n\mathcal V^2_n$ (or $n \mathcal
V_n^2/T_2$) are significant. The dCov test and test statistics are
implemented in the \emph{energy} package for R in functions \emph
{dcov.test}, \emph{dcov}, and \emph{dcor} \cite{R,energy}.

We have also obtained a result that gives an asymptotic critical value
applicable to arbitrary distributions. If $Q$ is a quadratic form of
centered Gaussian random variables and $E[Q]=1$, then
\[
P\{Q \ge\chi^2_{1-\alpha}(1)\} \le\alpha
\]
for all $0<\alpha\leq0.215$, where $\chi^2_{1-\alpha}(1)$ is the
$(1-\alpha)$ quantile of a chi-square variable with 1 degree of freedom.
This result follows from a theorem of Sz\'{e}kely and Bakirov
\cite{sb}, page~181.

Thus, a test that rejects independence if ${n \mathcal V^2_n / T_2} \ge
\chi^2_{1-\alpha}(1)$ has an asymptotic significance level at most $
\alpha$. This test criterion could be quite conservative for many
distributions.
Although this critical value is conservative, it is a sharp bound; the
upper bound $\alpha$ is achieved when $X$ and $Y$ are independent
Bernoulli variables.

\subsubsection*{Results for the bivariate normal distribution\label
{special}} When $(X,Y)$ has a bivariate normal distribution, there is a
deterministic relation between $\mathcal R$ and $|\rho|$.

\begin{theorem} \label{T:abvn} If $X$ and $Y$ are standard normal,
with correlation $\rho= \rho(X,Y)$, then:
\begin{longlist}[(iii)]
\item[(i)] $ \mathcal{R}(X,Y)\le|\rho|, $

\item[(ii)] $
\mathcal{R}^2(X,Y)= {\frac{\rho\arcsin\rho+\sqrt{1-\rho
^2}-\rho\arcsin({\rho}/2)-\sqrt{4-\rho^2}+1 }{1+\pi/3-\sqrt{3}}},
$
\item[(iii)] $\inf_{\rho\neq0} \frac{\mathcal{R}(X,Y)}{|\rho|} =
\lim_{\rho\to0}
\frac{\mathcal{R}(X,Y)}{|\rho|}=\frac{1}{2(1+\pi/3-\sqrt
{3})^{1/2}} \approxeq
0.89066.$
\end{longlist}
\end{theorem}

\begin{figure}[b]

\includegraphics{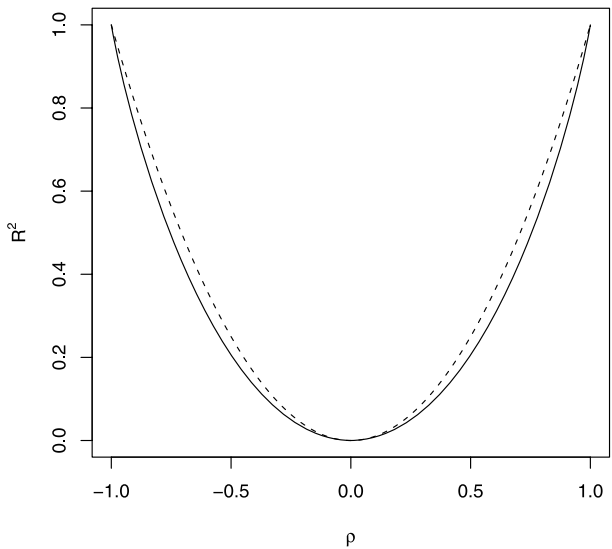}

 \caption{Dependence coefficient $\mathcal{R}^2$ (solid line) and
 correlation $\rho^2$ (dashed line) in the bivariate normal case.}
 \label{F:avsrho}
\end{figure}

The relation between $\mathcal R$ and $\rho$ for a bivariate normal
distribution is shown in Figure \ref{F:avsrho}.

\section{Brownian covariance}\label{sec3}
To introduce the notion of Brownian covariance, let us begin by
considering the squared product-moment covariance. Recall that a primed
variable $X'$ denotes an i.i.d. copy of the unprimed symbol $X$.
For two real-valued random variables, the square of their classical
covariance is
\begin{eqnarray}\label{cc}
&&E^2\bigl[\bigl(X-E(X)\bigr)\bigl(Y-E(Y)\bigr)\bigr]\nonumber
\\[-8pt]\\[-8pt]
&&\qquad = E\bigl[\bigl(X-E(X)\bigr)\bigl(X'-E(X')\bigr)\bigl(Y-E(Y)\bigr)\bigl(Y'-E(Y')\bigr)\bigr]. \nonumber
\end{eqnarray}
Now we generalize the squared covariance and define the square of
conditional covariance, given two real-valued stochastic processes
$U(\cdot)$ and $V(\cdot)$. We obtain an interesting result when $U$
and $V$ are independent Weiner processes.

First, to center the random variable $X$ in the conditional covariance,
we need the following definition. Let $X$ be a real-valued random
variable and $\{U(t)\dvtx t \in\mathbb R^1\}$ a real-valued stochastic
process, independent of $X$. The $U$-centered version of $X$ is defined by
%
\begin{equation}\label{central}
X_U = U(X)-\int_{-\infty}^\infty U(t)\, dF_X(t) = U(X) - E[U(X)
\vert U],
\end{equation}
whenever the conditional expectation exists.

Note that if $\mathit{id}$ is identity, we have $X_{\mathit{id}}=X-E[X]$. The important
examples in this paper apply Brownian motion/Weiner processes.

\subsection{Definition of Brownian covariance}

Let $W$ be a two-sided one-dimen\-sional Brownian motion/Wiener
process with expectation zero and covariance function
%
\begin{equation}\label{e:covfn}
|s| + |t| - |s-t|= 2\min(s,t), \qquad t,s\ge0.
\end{equation}
This is twice the covariance of the standard Wiener process. Here
the factor 2 simplifies the computations, so throughout the paper,
covariance function (\ref{e:covfn}) is assumed for $W$.

\begin{definition}
The Brownian covariance or the  Wiener
covariance of two real-valued random variables $X$ and $Y$ with finite
second moments
is a non-negative number defined by its square
%
\begin{equation}\label{defCovW}
\mathcal{W}^2(X,Y) = \operatorname{Cov}^2_W (X,Y)= E[X_W X'_W Y_{W'}Y'_{W'}],
\end{equation}
where $(W,W')$ does not depend on $(X, Y, X',Y')$.
\end{definition}

Note that if $W$ in $\operatorname{Cov}_W$ is replaced by the (nonrandom) identity
function~$\mathit{id}$, then
$\operatorname{Cov}_{\mathit{id}} (X,Y) = |\operatorname{Cov}(X,Y)| = |\sigma_{X,Y}|$, the absolute value
of Pearson's product-moment covariance.
While the standardized product-moment covariance, Pearson
correlation ($\rho$), measures the degree of linear
relationship between two real-valued variables, we shall see that standardized
Brownian covariance measures the degree of {\it all kinds of
possible relationships} between two real-valued random variables.

The definition of $\operatorname{Cov}_W(X,Y)$ can be extended to random processes in
higher dimensions as follows. If $X$ is an $\mathbb R^p$-valued random
variable, and $U(s)$ is a random process (random field) defined for all
$s \in\mathbb R^p$ and independent of $X$, define the $U$-centered
version of $X$ by
\[
X_U = U(X) - E[U(X) \vert U],
\]
whenever the conditional expectation exists.

\begin{definition}
If $X$ is an $\mathbb R^p$-valued random variable, $Y$ is an $\mathbb
R^q$-valued
random variable, and $U(s)$ and $V(t)$ are arbitrary random processes
(random fields) defined for all $s \in\mathbb R^p$, $t \in\mathbb
R^q$, then the $(U,V)$ covariance of $(X,Y)$ is defined as the
nonnegative number
whose square is
%
\begin{equation}\label{CovUV1}
\operatorname{Cov}^2_{U,V}(X,Y)=E[X_U X'_U Y_V Y'_V],
\end{equation}
whenever the right-hand side is nonnegative and finite.

In particular, if $W$ and $W'$ are independent Brownian motions with
covariance function (\ref{e:covfn}) on $\mathbb R^p$, and $\mathbb
R^q$ respectively, the \emph{Brownian covariance} of $X$ and
$Y$ is defined by
%
\begin{equation}\label{defBCpq}
\mathcal{W}^2 (X,Y) =\operatorname{Cov}^2_W(X,Y)=\operatorname{Cov}^2_{W, W'}(X,Y).
\end{equation}

Similarly, for random variables with finite variance define the
\emph{Brownian variance} by
\[
\mathcal{W}(X) = \operatorname{Var}_W(X) = \operatorname{Cov}_W(X,X).
\]
\end{definition}

\begin{definition}
The Brownian correlation is defined as
\[
\operatorname{Cor}_W(X,Y)
= \frac{\mathcal{W} (X,Y)}{ \sqrt{ \mathcal W (X) \mathcal{W} (Y)}}
\]
whenever the denominator is not zero; otherwise $\operatorname{Cor}_W (X,Y)=0$.
\end{definition}

In the following sections we prove that $\operatorname{Cov}_W(X,Y)$ exists for
random vectors~$X$ and $Y$ with finite second moments, and derive the
Brownian covariance in this case.

\subsection{Existence of $\mathcal W(X,Y)$}

In the following, the subscript on Euclidean norm $|x|_d$ for $x \in
\mathbb R^d$ is omitted when the dimension is self-evident.

\begin{theorem}\label{T1}If $X$ is an $\mathbb R^p$-valued
random variable, $Y$ is an $\mathbb R^q$-valued random variable, and
$E(|X|^2 + |Y|^2) < \infty$, then $ E[X_WX'_WY_{W'}Y_{W'}']$ is
nonnegative and finite, and
%
\begin{eqnarray}\label{e:T1}
\mathcal{W}^2(X,Y) &=& E[X_WX'_WY_{W'}Y_{W'}']\nonumber
\\
&=&
E|X-X'|| Y-Y'| + E|X-X'|E|Y-Y'|
\\
&&{}- E|X-X'|| Y-Y''| - E|X-X''|| Y-Y'|, \nonumber
\end{eqnarray}
where $(X,Y)$, $(X',Y')$, and $(X'', Y'')$ are i.i.d.
\end{theorem}

\begin{pf}
Observe that
\begin{eqnarray*}
E[ X_W X'_W Y_{W'}Y_{W'}' ]&=& E [ E(X_W Y_{W'}X'_W
Y_{W'}'|W,W') ]
\\
&=& E[ E(X_WY_{W'}|
W, W') E(X'_W Y_{W'}'| W, W') ]
 \\
 &=& E
[ E(X_WY_{W'}| W,W') ]^2,
\end{eqnarray*}
and this is always nonnegative. For finiteness, it is enough to prove
that all factors in the definition of $\operatorname{Cov}_W(X,Y)$ have finite fourth
moments. Equation (\ref{e:T1}) relies on the special form of the
covariance function (\ref{e:covfn}) of $W$. The remaining details are
in the \hyperref[app]{Appendix}.
\end{pf}

See Section \ref{alpha} for definitions and extension of results for
the general case of fractional Brownian motion with Hurst parameter
$0<H<1$ and covariance function $|t|^{2H}+|s|^{2H}-|t-s|^{2H}$.

\subsection{The surprising coincidence: $\mathcal{W} = \mathcal{V}$}
\label{S4}

\begin{theorem} \label{Th8}
For arbitrary $X\in\mathbb R^p$, $Y\in\mathbb R^q$ with finite second moments
\[
\mathcal{W}(X,Y)= \mathcal{V}(X,Y).
\]
\end{theorem}

\begin{pf}
Both $\mathcal{V}$ and $\mathcal{W}$ are nonnegative, hence, it is
enough to show that their squares coincide. Lemma \ref{lemmaC} can be
applied to evaluate $\mathcal V^2(X,Y)$.
In the numerator of the integral we have terms like
\[
E[ \cos\langle X-X',t \rangle
\cos\langle Y-Y',s \rangle],
\]
where $X,X'$ are i.i.d. and $Y,Y'$ are i.i.d. Now apply the identity
\[
\cos u \cos v
= 1 - (1 - \cos u) - (1 - \cos v) + (1-\cos u)(1-\cos v)
\]
and Lemma \ref{lemmaC} to simplify the integrand.
After cancelation in the numerator of the integrand, there remains to
evaluate integrals of the type
\begin{eqnarray*}
&& E\int_{\mathbb{R}^{p+q}} \frac{[1-\cos\langle X-X',t \rangle
][1-\cos\langle
Y-Y',s \rangle)]}{|t|^{1+p}
|s|^{1+q}} \,dt\, ds
\\
&&\qquad =
E \biggl[ \int_{\mathbb{R}^{p}} \frac{1-\cos\langle X-X',t \rangle
}{|t|^{1+p}}\, dt
\times
\int_{\mathbb{R}^{q}} \frac{1-\cos\langle Y-Y',s \rangle
}{|s|^{1+q}}\, ds
\biggr]
\\
&&\qquad = c_pc_q E|X-X'| E|Y-Y'|.
\end{eqnarray*}
Applying similar steps, after further simplification, we obtain
\begin{eqnarray*}
\mathcal{V}^2(X,Y) &=& E|X-X'||Y-Y'| + E|X-X'|E|Y-Y'|
\\
 &&{}
- E|X-X'|| Y- Y''| - E|X - X''|| Y - Y'|,
\end{eqnarray*}
and this is exactly equal to the expression (\ref{e:T1}) obtained for
$\mathcal W(X,Y)$ in Theorem~\ref{T1}.
\end{pf}

As a corollary to Theorem \ref{Th8}, the properties of Brownian
covariance for random vectors $X$ and $Y$ with finite second moments are
therefore the same properties established for distance covariance
$\mathcal V(X,Y)$ in Theorem \ref{Th3}.

The surprising result that Brownian covariance equals distance
covariance $\operatorname{dCov}$, exactly as defined in (\ref{Adef})
for $X \in
\mathbb R^p$ and $Y \in\mathbb R^q$, parallels a familiar special case when
$p=q=1$. For bivariate $(X,Y)$ we found that $\mathcal R(X,Y)$ is a
natural counterpart of the \emph{absolute value of} the Pearson correlation.
That is, if in (\ref{CovUV1}) $U$ and $V$ are the simplest nonrandom function
$\mathit{id}$, then we obtain the square of Pearson covariance $\sigma
_{X,Y}^2$. Next, if
we consider the most fundamental random processes, $U=W$ and $V=W'$, we
obtain the square of distance covariance, $\mathcal
V^2(X,Y)$.\looseness=1

Interested readers are referred to Sz\'{e}kely and Bakirov \cite{TR08} for
the background of the interesting coincidence in Theorem \ref{Th8}.

\section{Extensions}\label{sec4}

\subsection{The class of $\alpha$-distance dependence measures}\label{alpha}

In two contexts above we have introduced dependence measures based on
Euclidean distance and on Brownian motion with Hurst index $H=1/2$
(self-similarity index). Our definitions and results can be extended to
a one-parameter family of distance dependence measures indexed by a
positive exponent $0<\alpha<2$ on Euclidean distance, or
equivalently by an index $h$, where $h=2H$ for Hurst parameters $0 < H
< 1$.\looseness=1

If $E(|X|^\alpha_p+|Y|^\alpha_q)<\infty$ define $\mathcal
V^{(\alpha)}$ by its square
\begin{eqnarray*}
\mathcal{V}^{2(\alpha)}(X,Y) &=&
\| f_{X,Y}(t,s)-
f_{X}(t)f_{Y}(s)\|^2_\alpha
\\
&=& \frac{1}{C({p,\alpha})C({q,\alpha})}
\int_{\mathbb{R}^{p+q}}
\frac{| f_{X,Y}(t,s)-
f_{X}(t)f_{Y}(s)|^2} {|t|_p^{\alpha+p} |s|_q^{\alpha+q}} \,dt \,ds.
\end{eqnarray*}
Similarly, $\mathcal R^{(\alpha)}$ is the square root of
\[
\mathcal R^{2(\alpha)} = \frac{\mathcal V^{2(\alpha)}(X,Y)}
{\sqrt{\mathcal V^{2(\alpha)}(X)\mathcal V^{2(\alpha)}(Y)}}, \qquad
0 < \mathcal V^{2(\alpha)}(X), \mathcal V^{2(\alpha)}(Y) <
\infty,
\]
and $\mathcal R^{(\alpha)}=0$ if $\mathcal V^{2(\alpha)}(X) \mathcal
V^{2(\alpha)}(Y)=0$.

Now consider the L\'{e}vy fractional Brownian motion $\{W_H^d(t), t\in
\mathbb R^d\}$ with Hurst index $H\in(0,1)$, which
is a centered Gaussian random process with covariance function
\[
E[W_H^d(t)W_H^d(s)]=|t|^{2H}+|s|^{2H}-|t-s|^{2H},\qquad t,s\in\mathbb R^d.
\]
See Herbin and Merzbach \cite{hm07}.

In the following, $(W_H,W'_{H^*})$ and $(X,X',Y,Y')$ are supposed to be
independent.\looseness=1

Using Lemma \ref{lemmaC}, it can be shown for Hurst parameters $ 0<
H$, $H^* \le1$, $h:=2H$, and $h^*:=2H^*$, that
\begin{eqnarray}\label{e:hexp}
&&\operatorname{Cov}^2_{ W^p_{H}, W^{'q}_{H^*}}  (X,Y)\nonumber
\\
&&\qquad =
\frac1{C(p,h)C(q,h^*)}\int_{\mathbb R^p} \int
_{\mathbb R^q}\frac{|f(t,s)-f(t)g(s)|^2
\,dt\, ds}{|t|_p^{p+h} |s|_q^{q+ h^*}}\nonumber
\\[-8pt]\\[-8pt]
&&\qquad =
E|X-X'|_p^{h}|Y-Y'|_q^{h^*}+E|X-X'|_p^{h}E|Y-Y'|_q^{h^*}\nonumber
\\
&&\qquad\quad {} -E|X-X'|_p^{h}|Y-Y''|_q^{h^*} - E|X-X''|_p^{h}|Y-Y'|_q^{h^*}.\nonumber
\end{eqnarray}
Here we need to suppose that $E|X|_p^{2h}<\infty$,
$E|Y|_q^{2h^*}<\infty$.
Observe that when $h=h^*=1$, (\ref{e:hexp}) is equation (\ref{e:T1})
of Theorem \ref{T1}.

The corresponding statistics are defined by replacing
the exponent 1 with exponent $\alpha$ (or $h$) in the distance dependence
statistics (\ref{e:AnXY}), (\ref{e:AnX}), and (\ref{e:alphan}). That
is, in the sample distance matrices replace $a_{kl} = |X_k - X_l|_p$ with
$a_{kl} = |X_k - X_l|_p^\alpha,$ and replace $b_{kl} = |Y_k- Y_l|_q$
with $b_{kl} = |Y_k- Y_l|_q^\alpha$, $k,l=1,\dots,n$.

Theorem \ref{Thm2} can be generalized for $\| \cdot\|_\alpha$
norms, so that almost sure convergence of $\mathcal V_n^{(\alpha)}
\to\mathcal V^{(\alpha)}$ follows if the $\alpha$-moments are
finite. Similarly, one can prove the weak convergence and statistical
consistency for $\alpha$ exponents, $0 < \alpha< 2$, provided that
$\alpha$ moments are finite.

Note that the strict inequality $0<\alpha<2$ is important. Although
$\mathcal V^{(2)}$ can be defined for $\alpha=2$, it does not
characterize independence.
Indeed, the case $\alpha=2$ (squared Euclidean distance) leads to
classical product-moment correlation and covariance for bivariate
$(X,Y)$. Specifically, if $p=q=1$, then $\mathcal R^{(2)}=|\rho|$,
$\mathcal R_n^{(2)}=|\hat\rho|$, and $\mathcal
V_n^{(2)}=2|\hat\sigma_{xy}|$, where $\hat\sigma_{xy}$ is the
maximum likelihood estimator of Pearson covariance $\sigma
_{x,y}=\sigma(X,Y)$.

\subsection{Affine invariance}\label{S:affine}
Independence is preserved under affine transformations hence it is
natural to consider dependence measures that are affine invariant. We
have seen that $\mathcal{R}(X,Y)$ is invariant with respect to
orthogonal transformations
%
\begin{equation}\label{e:affine}
X \mapsto a_1 + b_1 C_1X, \qquad Y \mapsto a_2 + b_2 C_2 Y,
\end{equation}
where $a_1$, $a_2$ are arbitrary vectors, $b_1$, $b_2$ are arbitrary
nonzero numbers, and $C_1$, $C_2$ are arbitrary orthogonal
matrices. We can also define a distance correlation that is affine
invariant.
Define the scaled samples $\mathbf{X}^*$ and
$\mathbf{Y}^*$ by
%
\begin{equation}\label{e:xystar}
\mathbf{X}^*=\mathbf{X} {S}_X^{-1/2}, \qquad
\mathbf{Y}^*=\mathbf{Y} {S}_Y^{-1/2},
\end{equation}
where ${S}_X$ and ${S}_Y$ are the sample covariance matrices of
$\mathbf{X}$ and $\mathbf{Y}$ respectively. The sample
vectors in (\ref{e:xystar}) are not invariant to affine
transformations, but the distances, $|X^*_k-X^*_l|$ and $|Y^*_k-Y^*_l|$,
$k,l=1,\dots,n$, are invariant to affine transformations. Thus, an affine
distance correlation statistic can be defined by its square
\[
\mathcal{R}_n^{*2}(\mathbf{X,Y}) =
\frac{\mathcal{V}^2_n(\mathbf{X}^*,\mathbf{Y}^*)}
{\sqrt{\mathcal{V}^2_n(\mathbf{X}^*)
\mathcal{V}^2_n(\mathbf{Y}^*)}}.
\]

Theoretical properties established for $\mathcal V_n$ and $\mathcal
R_n$ also hold for
$\mathcal{V}^*_n$ and $\mathcal{R}^*_n$, because the transformation
simply replaces the original weight function
$\{c_p c_q |t|_p^{1+p} |s|_q^{1+q}\}^{-1}$ with $
\{c_p c_q |\Sigma_X^{1/2} t|_p^{1+p} |\Sigma_Y^{1/2}
s|_q^{1+q}\}^{-1}.
$

\subsection{Rank test}

In the case of bivariate $(X,Y)$ one can also consider a distance
covariance test of independence for rank($X$), rank($Y$), which has the
advantage that it is distribution free and invariant with respect to
monotone transformations of $X$ and $Y$, but usually at a cost of lower
power than the $\operatorname{dCov}(X,Y)$ test (see Example \ref
{exNIST}). The
rank-dCov test can be applied to continuous or discrete data, but for
discrete data it is necessary to use the correct method for breaking
ties. Any ties in ranks should be broken randomly, so that a sample of
size $n$ is transformed to some permutation of the integers 1:$n$. A
table of critical values for the statistic $n \mathcal R_n^2$, based on
Monte Carlo results, is provided in Table \ref{cvs} in the \hyperref[app]{Appendix}.

\section{Applications}\label{sec5}

\subsection{Nonlinear and nonmonotone dependence}

Suppose that one wants to test the independence of $X$ and $Y$, where
$X$ and $Y$ cannot be observed directly, but can only be measured with
independent errors. Consider the following:
\begin{longlist}[(iii)]
\item[(i)] Suppose that $X_i$ can only be measured through observation of
$A_i = X_i+ \varepsilon_i$, where $\varepsilon_i$ are independent of
$X_i$, and similarly for $Y_i$.
\item[(ii)] One can only measure (non) random functions of $X$ and $Y$, for
example,
$A_i = \phi(X_i)$ and $B_i=\psi(Y_i)$.
\item[(iii)] Suppose both (i) and (ii) for certain types
of random $\phi$ and $\psi$.
\end{longlist}
In all of these cases, even if $(X,Y)$ were jointly normal, the dependence
between $(A,B)$ can be such that the correlation of $A$ and $B$ is
almost irrelevant, but $\operatorname{dCor}(A,B)$ is obviously relevant.

In this section we illustrate a few of the many possible applications
of distance covariance. The dCov test has been applied using the \emph
{dcov.test} function in the \emph{energy} \cite{energy} package for R
\cite{R}, where it is implemented as a permutation test.

\begin{figure}[t]

\includegraphics{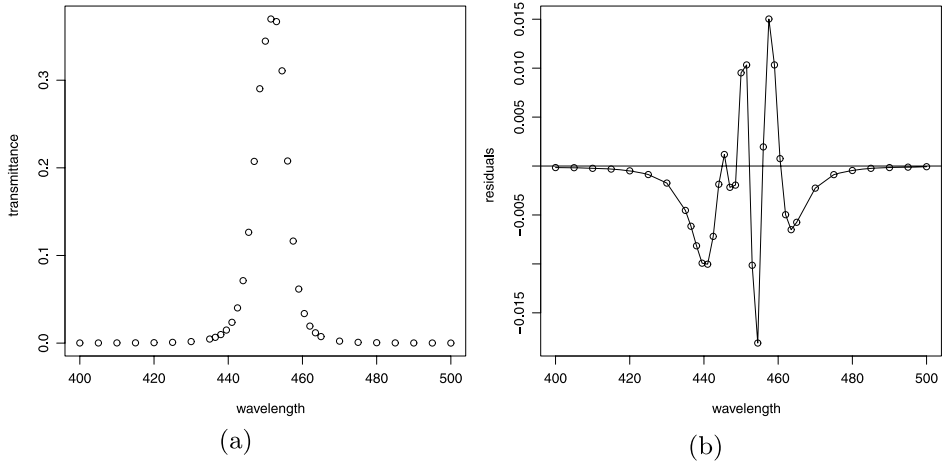}

 \caption{The Eckerle4 data \textup{(a)} and plot of residuals vs predictor
 variable for the NIST certified estimates \textup{(b)}, in Example \protect\ref{exNIST}.}
 \label{fig2}
\end{figure}

\subsection{Examples}

\begin{example}\label{exNIST}
This example is similar to the type
considered in (ii), with observed data from the NIST Statistical
Reference Datasets (NIST StRD) for Nonlinear Regression. The data
analyzed is \emph{Eckerle4}, data from an NIST study of circular
interference transmittance \cite{eck79}. There are 35 observations,
the response variable is transmittance, and the predictor variable is
wavelength. A plot of the data in Figure \ref{fig2}(a) reveals that
there is a nonlinear relation between wavelength and transmittance. The
proposed nonlinear model is
\[
y = f(x;\beta)+\varepsilon= \frac{\beta_1}{\beta_2}
\exp\biggl\{ \frac{-(x-\beta_3)^2}{2\beta_2^2} \biggr\} +
\varepsilon,
\]
where $\beta_1, \beta_2 >0$, $\beta_3\in\mathbb R$, and
$\varepsilon$ is random error.
In the hypothesized model, $Y$ depends on the density of $X$.

Results of the dCov test of independence of wavelength and
transmittance are
{

\footnotesize{
\begin{verbatim}

         dCov test of independence
  data: x and y
  nV^2 = 8.1337, p-value = 0.021
  sample estimates:
       dCor
  0.4275431

\end{verbatim}}}
\noindent with $ {\mathcal R}_n \doteq0.43$, and dCov is significant ($p$-value
$=$
0.021) based on 999 replicates. In contrast, neither Pearson
correlation $\hat\rho=0.0356$, ($p$-value $=$ 0.839) nor Spearman rank
correlation $\hat\rho_s= 0.0062$ ($p$-value $=$ 0.9718) detects the
nonlinear dependence between wavelength and transmittance, even though
the relation in Figure \ref{fig2}(a) appears to be nearly deterministic.

The certified estimates (best solution found) for the parameters are
reported by NIST as $\hat\beta_1\doteq1.55438$, $\hat\beta_2
\doteq4.08883$, and $\hat\beta_3 \doteq451.541$. The residuals of
the fitted model are easiest to analyze when plotted vs the predictor
variable as in Figure \ref{fig2}(b). Comparing residuals and transmittance,
{

\footnotesize{
\begin{verbatim}
         dCov test of independence
  data: y and res
  nV^2 = 0.0019, p-value = 0.019
  sample estimates:
       dCor
  0.4285534

\end{verbatim}}}
\noindent
we have $\mathcal R_n \doteq0.43$ and the dCov test is significant
($p$-value = 0.019) based on 999 replicates.
Again the Pearson correlation is nonsignificant ($\hat\rho\doteq
0.11$, $p$-value~$=$ 0.5378).

Although nonlinear dependence is clearly evident in both plots, note
that the methodology applies to multivariate analysis as well, for
which residual plots are much less informative.
\end{example}

\begin{figure}[b]

\includegraphics{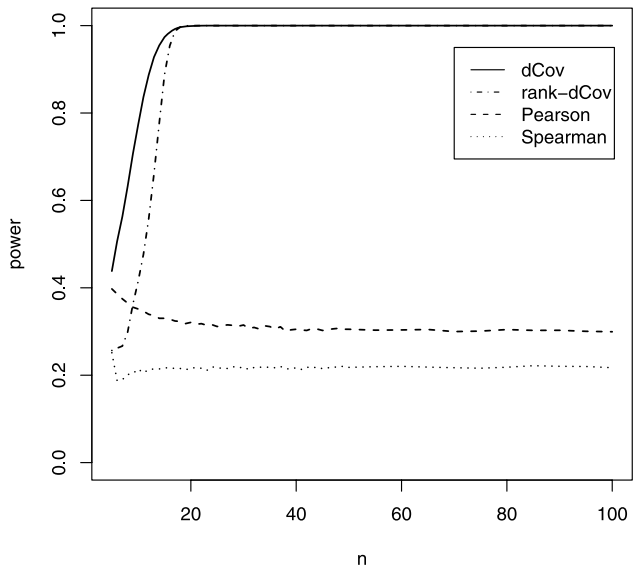}

 \caption{Example \protect\ref{exPower}: Empirical power at $0.1$
 significance and sample size $n$.}\label{F:power}
\end{figure}

\begin{example}In the model specification of Example \ref{exNIST}, the
response variable $Y$ is assumed to be proportional to a normal density
plus random error. For simplicity, consider $(X,Y)=(X,\phi(X))$, where
$X$ is standard normal and $\phi(\cdot)$ is the standard normal
density. Results of a Monte Carlo power comparison of the dCov test
with classical Pearson correlation and Spearman rank tests are shown in
Figure \ref{F:power}. The power estimates are computed as the
proportion of significant tests out of 10,000 at 10\% significance level.

In this example, where the relation between $X$ and $Y$ is
deterministic but not monotone, it is clear that the dCov test is
superior to product moment correlation tests. Statistical consistency
of the dCov test is evident, as its power increases to~1 with sample
size, while the power of correlation tests against this alternative
remains approximately level across sample sizes. We also note that
distance correlation applied to ranks of the data is more powerful in
this example than either correlation test, although somewhat less
powerful than the dCov test on the original $(X,Y)$ data. \label{exPower}
\end{example}

\begin{figure}[b]

\includegraphics{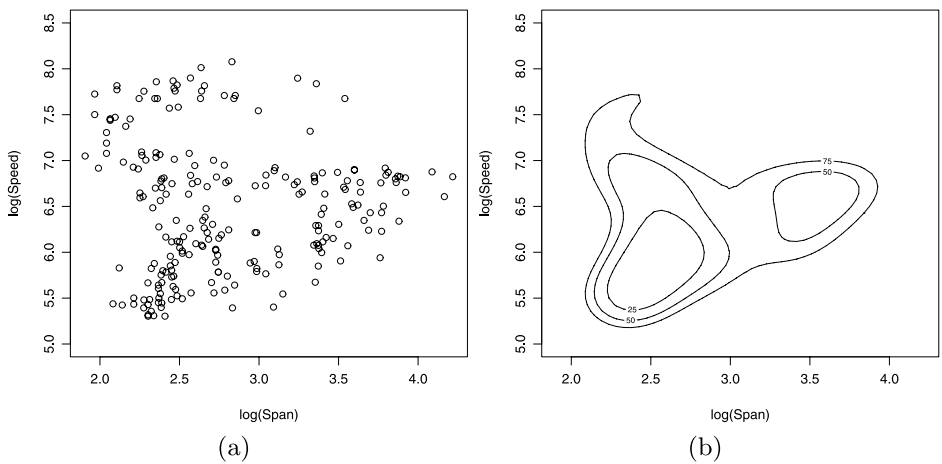}

 \caption{Scatterplot and contours of density estimate for the aircraft speed and span variables, period 3,
 in Example \protect\ref{exAir}.}\label{fig4}
\end{figure}

\begin{example}\label{exAir}
The Saviotti aircraft data \cite{sav96} record six characteristics of
aircraft designs which appeared during the twentieth century. We
consider two variables, wing span (m) and speed (km/h) for the 230
designs of the third (of three) periods. This example and the data
(\emph{aircraft}) are from Bowman and Azzalini \cite{ba97,sm}. A
scatterplot on log-log scale of the variables and contours of a
nonparametric density estimate are shown in Figures \ref{fig4}(a)
and \ref{fig4}(b). The nonlinear relation between speed and wing
span is quite evident from the plots.

The dCov test of independence of log(Speed) and log(Span) in period 3
is significant ($p$-value $=$ 0.001), while the Pearson correlation test is
not significant ($p$-value $=$ 0.8001).
{

\footnotesize{
\begin{verbatim}

         dCov test of independence
  data: logSpeed3 and logSpan3
  nV^2 = 3.4151, p-value = 0.001
  sample estimates:
       dCor
  0.2804530

         Pearson's product-moment correlation
  data: logSpeed3 and logSpan3
  t = 0.2535, df = 228, p-value = 0.8001
  alternative hypothesis: true correlation is not equal to 0
  95 percent confidence interval:
  -0.1128179 0.1458274
  sample estimates:
         cor
  0.01678556

\end{verbatim}}}
\noindent
The sample estimates are $\hat\rho= 0.0168$ and $\mathcal R_n =
0.2805$. Here we have an example of observed data where two variables
are nearly uncorrelated, but dependent. We obtained essentially the
same results on the correlations of ranks of the data.
\end{example}

\begin{example}\label{exCrime}
This example compares dCor and Pearson correlation in exploratory data
analysis. Consider the Freedman \cite{census70,freedman} data on crime
rates in~US metropolitan areas with 1968 populations of 250,000 or
more. The data set is available from Fox \cite{car}, and contains four
numeric variables:

\begin{itemize}
\item[\textit{population}] (total 1968, in thousands),
\item[\textit{nonwhite}] (percent nonwhite population, 1960),
\item[\textit{density}] (population per square mile, 1968),
\item[\textit{crime}] (crime rate per 100,000, 1969).
\end{itemize}

The 110 observations contain missing values. The data analyzed are the
100 cities with complete data. Pearson $\hat\rho$ and dCor statistics
$\mathcal R_n$ are shown in Table \ref{T:Freedman}. Note that there is
a significant association between crime and population density measured
by dCor, which is not significant when measured by $\hat\rho$.

Analysis of this data continues in Example \ref{exJack}.
\end{example}

\begin{table}[b]
\caption{Pearson correlation and distance correlation statistics for
the Freedman data of Example \protect\ref{exCrime}. Significance at
$0.05,0.01,0.001$ for the corresponding tests is indicated by
$*,**,***$, respectively}\label{T:Freedman}
\begin{tabular*}{\tablewidth}{@{\extracolsep{4in minus 4in}}lcccccc@{}}
\hline
&\multicolumn{3}{c}{\textbf{Pearson}}& \multicolumn{3}{c@{}}{\textbf{dCor}} \\[-6pt]
&\multicolumn{3}{c}{\hrulefill}& \multicolumn{3}{c@{}}{\hrulefill} \\
& \textbf{Nonwhite} & \textbf{Density} & \textbf{Crime} & \textbf{Nonwhite} & \textbf{Density} & \textbf{Crime} \\
\hline
Population & 0.070 & 0.368$^{***}$ & 0.396$^{***}$ & 0.260$^{*}$ & 0.615$^{***}$ & 0.422$^{**}$\phantom{$^{*}$}
\\
Nonwhite & & 0.002\phantom{$^{***}$} & 0.294$^{**}$\phantom{$^{*}$} & & 0.194\phantom{$^{***}$} & 0.385$^{***}$ \\
Density & & & 0.112\phantom{$^{***}$} & & & 0.250$^{*}$\phantom{$^{**}$} \\
\hline
\end{tabular*}
\end{table}

\begin{figure}[b]

\includegraphics{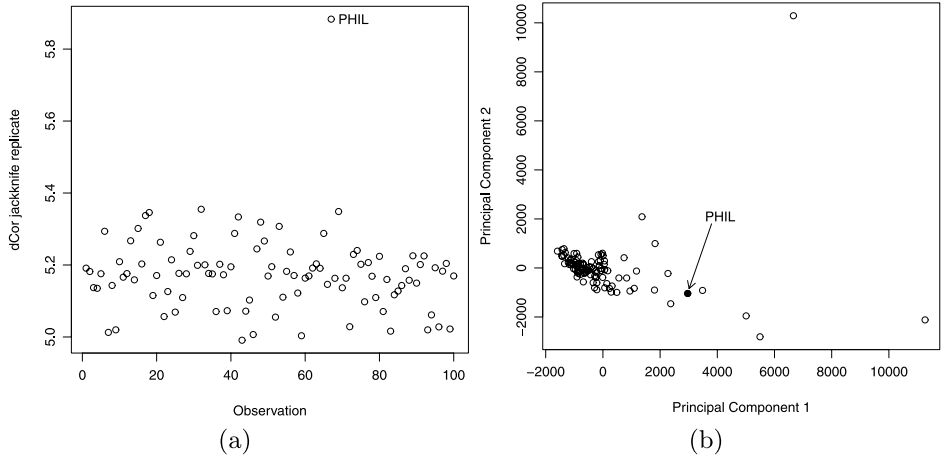}

 \caption{Jackknife replicates of dCor \textup{(a)} and principal components of Freedman
 data \textup{(b)} in Example~\protect\ref{exJack}.}\label{fig5}
\end{figure}

\begin{example}[(Influential observations)]\label{exJack}
When $\mathcal V_n$ and $\mathcal R_n$ are computed using formula (\ref
{e:AnXY}), it is straightforward to apply a jackknife procedure to
identify possible influential observations or to estimate standard
error of $\mathcal V_n$ or $\mathcal R_n$. A `leave-one-out' sample
corresponds to $(n-1) \times(n-1)$ matrices $A_{(i)kl}$ and
$B_{(i)kl}$, where the subscript $(i)$ indicates that the $i$th
observation is left out. Then $A_{(i)kl}$ is computed from distance
matrix $A=(a_{kl})$ by omitting the $i$th row and the $i$th column of
$A$, and similarly $B_{(i)kl}$ is computed from $B=(b_{kl})$ by
omitting the $i$th row and the $i$th column of $B$. Then
\[
\mathcal V^2_{(i)}(\mathbf X, \mathbf Y) = \frac1 {(n-1)^2} \sum_{k,l
\neq i} A_{(i)kl}B_{(i)kl},
\qquad i=1,\ldots,n,
\]
are the jackknife replicates of $\mathcal V_n^2$, obtained without
recomputing matrices $A$ and $B$. Similarly, $\mathcal R^2_{(i)}$ can
be computed from the matrices $A$ and $B$. A jackknife estimate of the
standard error of $\mathcal R_n$ is thus easily obtained from the
matrices $A,B$ (on the jackknife, see, e.g., Efron and Tibshirani \cite{et93}).

The jackknife replicates $\mathcal R_{(i)}$ can be used to identify
potentially influential observations, in the sense that outliers within
the sample of replicates correspond to observations $X_i$ that increase
or decrease the dependence coefficient more than other observations.
These unusual replicates are not necessarily outliers in the original data.

Consider the crime data of Example \ref{exCrime}. The studentized
jackknife replicates $\mathcal R_{(i)}/{\widehat{se}( \mathcal
R_{(i)})}$, $i=1,\dots,n$, are plotted in Figure \ref{fig5}(a).
These replicates were computed on the pairs $(x,y)$, where $x$ is the
vector (\emph{nonwhite, density, population}) and $y$ is \emph
{crime}. The plot suggests that Philadelphia is an unusual observation.
For comparison we plot the first two principal components of the four
variables in Figure~\ref{fig5}(b), but Philadelphia (PHIL) does
not appear to be an unusual observation in this plot or other plots
(not shown), including those where log(\emph{population})
replaces
\emph{population} in the analysis. One can see from comparing
{

\footnotesize{
\begin{verbatim}

               population nonwhite density crime
  PHILADELPHIA       4829     15.7    1359  1753

\end{verbatim}}}
\noindent
with sample quartiles
{\footnotesize{
\begin{verbatim}

       population nonwhite density  crime
  0%       270.00    0.300     37.00   458.00
  25%      398.75    3.400    266.50  2100.25
  50%      664.00    7.300    412.00  2762.00
  75%     1167.75   14.825    773.25  3317.75
  100%   11551.00   64.300  13087.00  5441.00

\end{verbatim}}}
\noindent
that \emph{crime} in Philadelphia is low while \emph{population},
\emph{nonwhite}, and \emph{density} are all high relative to other
cities. Recall that all Pearson correlations were positive in Example
\ref{exCrime}.

This example illustrates that having a single multivariate summary
statistic dCor that measures dependence is a valuable tool in
exploratory data analysis, and it can provide information about
potential influential observations prior to model selection.
\end{example}

\begin{example}\label{exGumbel}In this example we illustrate how to
isolate the nonlinear dependence between random vectors to test for
nonlinearity.

Gumbel's bivariate exponential distribution \cite{gumbel61} has
density function
\[
f(x,y;\theta) = [(1+\theta x)(1 + \theta y)] \exp(-x-y-\theta xy),
\qquad x,y>0; 0 \leq\theta\leq1.
\]
The marginal distributions are standard exponential, so there is a
strong nonlinear, but monotone dependence relation between $X$ and $Y$.
The conditional density is
\[
f(y|x) = e^{-(1+\theta x)y}[(1+\theta x)(1+\theta y) - \theta],
\qquad y > 0.
\]
If $\theta= 0,$ then $f_{X,Y}(x,y)=f_{X}(x) f_{Y}(y)$ and independence
holds, so $\rho=0$. At the opposite extreme, if $\theta=1,$ then
$\rho= -0.40365$ (see Kotz, Balakrishnan, and Johnson \cite{CMD1},
Section~2.2). Simulated data was generated using the conditional
distribution function approach outlined in Johnson \cite{johnson87}.
Empirical power of dCov and correlation tests for the case $\theta
=0.5$ are compared in Figure \ref{fig6}(a), estimated from 10,000 test
decisions each for sample sizes \{10:100(10), 120:200(20), 250, 300\}.
This comparison reveals that the correlation test is more powerful than
dCov against this alternative, which is not unexpected because
$E[Y|X=x]=({1+\theta+x\theta})/{(1+x\theta)^2}$ is monotone.

While we cannot split the dCor or dCov coefficient into linear and
nonlinear components, we can extract correlation first and then compute
dCor on the residuals. In this way one can separately analyze the
linear and nonlinear components of bivariate or multivariate dependence
relations.

\begin{figure}[t]

\includegraphics{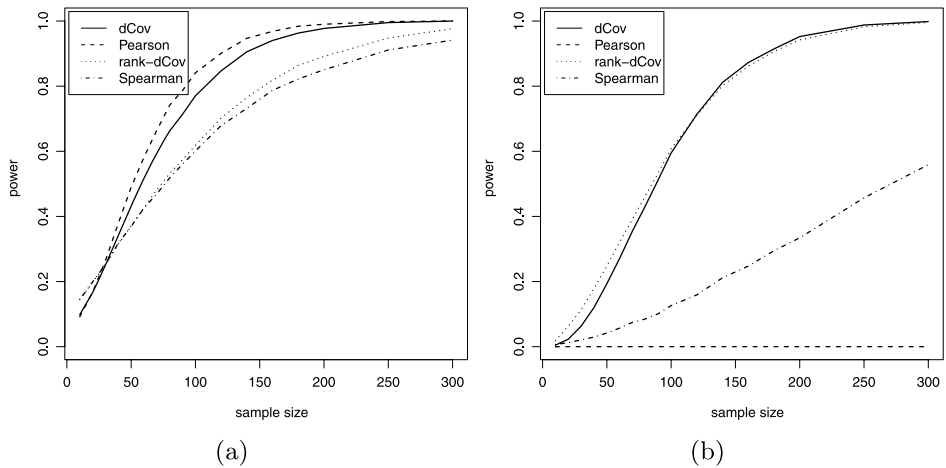}

 \caption{Power comparison of dCov and correlation
 tests at 10\% significance level for Gumbel's bivariate exponential distribution in Example
 \protect\ref{exGumbel}.}\label{fig6}
\end{figure}

To extract the linear component of dependence, fit a linear model
$Y=X\beta+\varepsilon$ to the sample $(\mathbf X, \mathbf Y)$ by
ordinary least squares. It is not necessary to test whether the linear
relation is significant. The residuals $\hat\varepsilon_i = X_i\hat
\beta- Y_i$ are uncorrelated with the predictors $\mathbf X$. Apply
the dCov test of independence to $(\mathbf X, \hat\varepsilon)$.

Returning to the Gumbel bivariate exponential example, we have
extracted the linear component and applied dCov to the residuals of a
simple linear regression model. Repeating the power comparison
described above on $(\mathbf X, \hat\varepsilon)$ data, we obtained
the power estimates shown in Figure \ref{fig6}(b). One can note
that power of dCov tests is increasing to 1 with sample size,
exhibiting statistical consistency against the nonlinear dependence
remaining in the residuals of the linear model.

This procedure is easily applied in arbitrary dimension. One can fit a
linear multiple regression model or a model with multivariate response
to extract the linear component of dependence. This has important
practical application for evaluating models in higher dimensions.
\end{example}

More examples, including Monte Carlo power comparisons for random
vectors in dimensions up to $p=q=30$, are given in Sz\'{e}kely et al.\
\cite{srb07}.

\section{Summary}\label{sec6}

Distance covariance and distance correlation are natural extensions and
generalizations of classical Pearson covariance and correlation in at
least two ways. In one direction we extend the ability to measure
linear association to all types of dependence relations. In another direction
we extend the bivariate measure to a single scalar measure of
dependence between random vectors in arbitrary dimension. In addition
to the obvious theoretical advantages, we have the practical advantages
that the dCov and dCor statistics are computationally simple, and
applicable in
arbitrary dimension not constrained by sample size.

We cannot claim that dCov is the only possible or the only reasonable
extension with the above mentioned properties, but we can claim that
our extension is a natural generalization of Pearson's covariance in
the following sense. We defined the covariance of random vectors with
respect to a pair of random processes, and if these random processes
are i.i.d. Brownian motions, which is a very natural choice, then we
arrive at the distance covariance; on the other hand, if we choose the
simplest nonrandom functions, a pair of identity functions (degenerate
random processes), then we arrive at Pearson's covariance.

We have illustrated only a few of the many applications where distance
correlation may provide additional information not measured by
classical correlation or arrays of bivariate statistics. In exploratory
data analysis, distance correlation has the flexibility to be applied
as a multivariate measure of dependence, or measure of dependence among
any of the lower dimensional marginal distributions.

The general linear model is fundamental in data analysis for several
reasons, but often a linear model is not adequate. We can test for
linearity using dCov as shown in Example \ref{exGumbel}. Although
illustrated for simple linear regression, the basic method is
applicable for all types of i.i.d. observations, including longitudinal
data or other data with multivariate predictors and/or multivariate response.

In summary, distance correlation is a valuable, practical, and natural
tool in data analysis and inference that extends the good properties of
classical correlation to multivariate analysis and the general
hypothesis of independence.

\appendix

\section{Proofs of statements}\label{app}

For $\mathbb R^d$ valued random variables, $|\cdot|_d$ denotes the
Euclidean norm; whenever the dimension is self-evident we suppress the
index $d$.

\subsection{\texorpdfstring{Proof of Theorem \protect\ref{Th3}(iii) and (vi)}{Proof of Theorem 3(iii) and (vi)}}\mbox{}

\begin{pf} Starting with the left side of the inequality (iii),
%
\begin{eqnarray}
&&\mathcal{V}(X_1+X_2,Y_1+Y_2)\nonumber
\\
&&\qquad =
\|f_{X_1+X_2,Y_1+Y_2}(t,s)-f_{X_1+X_2}(t)f_{Y_1+Y_2}(s) \|\nonumber
\\
&&\qquad=
\|f_{X_1,Y_1}(t,s)f_{X_2,Y_2}(t,s)-f_{X_1}(t)f_{X_2}(t)f_{Y_1}(s)f_{Y_2}(s) \nonumber
\|
\\ \label{iii1}
&&\qquad\leq
\bigl\|f_{X_1,Y_1}(t,s)\bigl( f_{X_2,Y_2}(t,s)-f_{X_2}(t)f_{Y_2}(s)
\bigr)\bigr\|
\\
 &&\qquad\quad{}+
\bigl\|f_{X_2}(t)f_{Y_2}(s) \bigl(
f_{X_1,Y_1}(t,s)-f_{X_1}(t)f_{Y_1}(s)\bigr)
\bigr\|\nonumber
\\
 \label{iii2} &&\qquad\leq
\| f_{X_2,Y_2}(t,s)-f_{X_2}(t)f_{Y_2}(s) \|
+ \|f_{X_1,Y_1}(t,s)-f_{X_1}(t)f_{Y_1}(s)\|
\\
 &&\qquad=
{\mathcal V}(X_1,Y_1)+{\mathcal V}(X_2,Y_2). \nonumber
\end{eqnarray}

It is clear that if (a) $X_1$ and $Y_1$ are both constants, (b) $X_2$
and $Y_2$ are both constants, or (c) $X_1, X_2, Y_1, Y_2$ are mutually
independent, then we have equality in (iii). Now suppose that we have
equality in (iii), and thus we have equality above at (\ref{iii1})
and (\ref{iii2}), but neither (a) nor (b) hold. Then the only way we
can have equality at (\ref{iii2}) is if $X_1 , Y_1$ are independent
and also $X_2, Y_2$ are independent. But our hypothesis assumes that
$(X_1,Y_1)$ and
$(X_2,Y_2)$ are independent hence (c) must hold.

Finally, (vi) follows from (iii). In this special case $X_1 = Y_1 = X$
and $X_2 = Y_2 = Y$. Now (a) means that $X$ is constant, (b) means that
$Y$ is constant, and (c) means that both of them are constants, because
this is the only case when a random variable can be independent of itself.
\end{pf}

\subsection{Existence of $\mathcal W(X,Y)$}

To complete the proof of Theorem \ref{T1}, we need to show that all
factors in the definition of $\operatorname{Cov}_W(X,Y)$ have finite fourth moments.

\begin{pf}
Note that $E[W^2(t)]=2|t|$, so that $E[W^4(t)]=3(
E[W^2(t)])^2=12|t|^2$ and, therefore,
\[
E[W^4(X)] =E[ E( W^4(X) | X ) ] =
E[12|X|^2] <\infty.
\]
On the other hand, by the inequality
$(a+b)^4\le2^4(a^4+b^4)$, and by Jensen's inequality, we have
\begin{eqnarray*}
E( X_W )^4 &=& E [ W(X)-E( W(X)| W
) ]^4
\\[2pt]
 &\le&
2^4\bigl( E[W^4(X)] + E [ E( W(X)|
W)]^4\bigr)
\\[2pt]
&\le& 2^5 E[W^4(X)]=2^5 12 E|X|^2 <
\infty.
\end{eqnarray*}
Similarly, the random variables $X_W'$, $Y_{W'}$, and $Y_{W'}'$ also
have finite fourth moments, hence,
\begin{eqnarray*}
\mathcal{W}^2(X,Y) &=& E [X_W X'_W Y_{W'}Y_{W'}']
\\[2pt]
&\le& \tfrac14 E[ ( X_W )^4+(
X'_W )^4 + ( Y_{W'})^4+( Y_{W'}' )^4
] <\infty.
\end{eqnarray*}

Above we implicitly used the fact that $E[W(X) \vert W]= \int
_{\mathbb R^p}
W(t)\, dF_X(t)$ exists a.s. This can easily be proved with the help of the
Borel--Cantelli lemma, using the fact that the supremum of centered
Gaussian processes have small tails (see
\cite{t88,ls70}).

Observe that
\begin{eqnarray*}
\mathcal{W}^2(X,Y) &=& E [X_W X_W' Y_{W'} Y_{W'}']
\\[2pt]
&=& E [ E( X_W X_W' Y_{W'}Y_{W'}' |
X,X',Y,Y') ]
\\[2pt]
&=& E [ E ( X_W X_W' | X,X',Y,Y') E(
Y_{W'}Y_{W'}' | X,X',Y,Y') ].
\end{eqnarray*}
Here
\begin{eqnarray*}
X_W X_W' &=& \biggl\{ W(X) - \int_{\mathbb R^p} W(t)\,dF_X(t) \biggr\}
\biggl\{W(X')- \int_{\mathbb R^p} W(t)\,dF_X(t) \biggr\}
\\
&=& W(X) W(X') - \int_{\mathbb R^p} W(X) W(t)\, dF_X(t)
\\
&&{} -
\int_{\mathbb R^p} W(X') W(t) \,dF_X(t)
+ \int_{\mathbb R^p}\int_{\mathbb R^p}W(t)W(s)\,dF_X(t)\,dF_X(s).
\end{eqnarray*}

By the definition of $W(\cdot)$, we have $ E [W(t)W(s)] = |t|+|s|-|t-s|,$
thus,
\begin{eqnarray*}
E [ X_W X_W'  | X,X',Y,Y' ] &=&
|X|+|X'|-|X-X'|
\\
&&{}-
\int_{\mathbb R^p}(|X|+|t|-|X-t|)\,dF_X(t)
\\&& {}-
\int_{\mathbb R^p}(|X'|+|t|-|X'-t|)\,dF_X(t)
\\
&& {}+ \int_{\mathbb R^p}
\int_{\mathbb R^p}(|t|+|s|-|t-s|)\,dF_X(t)\,dF_X(s).
\end{eqnarray*}
Hence,
\begin{eqnarray*}
E [ X_W X_W'  | X,X',Y,Y' ] &=&
|X|+|X'|-|X-X'|
\\
&& {}-(|X|+E|X|-E'|X-X'|)
\\
&& {}-(|X'|+E|X|-E''|X'-X''|)
\\
&& {}+
(E|X|+E|X'|-E|X-X'|)
\\
&=& E'|X-X'|+E''|X'-X''|-|X-X'|-E|X-X'|,
\end{eqnarray*}
where $E'$ denotes the expectation with respect to $X'$ and $E''$
denotes the expectation with respect to $X''$. A similar argument
for $Y$ completes the proof.
\end{pf}

\section{Critical values}\label{appb}

Estimated critical values for $n \mathcal R_n^2($rank($\mathbf{X}$),
rank($\mathbf{Y}$)) are summarized in Table \ref{cvs} for 5\% and
10\% significance levels. The critical values are estimates of the 95th
and 90th quantiles of the sampling distribution and were obtained by a
large scale Monte Carlo simulation (100,000 replicates for each $n$).
For sample sizes $n \leq10$, the probabilities were determined by
generating all possible permutations of the ranks, so the achieved
significance levels (ASL) reported for $n \leq10$ are exact. The
rejection region is in the upper tail.

\begin{table}[t]
\caption{Critical values of $n \mathcal
R_n^2($rank($\mathbf X),$ rank($\mathbf Y$)); exact achieved
significance level (ASL) for $n \leq10$, and Monte Carlo estimates for
$n \geq11$. Reject independence if $n \mathcal R_n^2$ is greater than
or equal to the~table~value}\label{cvs}
\begin{tabular*}{\tablewidth}{@{\extracolsep{4in minus 4in}}lcccccccc@{}}
\hline
$\bolds{n}$ & \textbf{10\% (ASL)} & \textbf{5\% (ASL)} & $\bolds{n}$ & \textbf{10\%} & \textbf{5\%} & $\bolds{n}$ & \textbf{10\%} & \textbf{5\%} \\
\hline
\phantom{0}5 & 3.685 (0.100) & 4.211 (0.050) & 15 & 4.25 & 5.16 & \phantom{0}25 & 4.26 & 5.22 \\
\phantom{0}6 & 3.917 (0.097) & 4.699 (0.047) & 16 & 4.25 & 5.17 & \phantom{0}30 & 4.25 & 5.22\\
\phantom{0}7 & 4.215 (0.098) & 4.858 (0.047) & 17 & 4.25 & 5.17 & \phantom{0}35 & 4.24 & 5.23 \\
\phantom{0}8 & 4.233 (0.099) & 4.995 (0.050) & 18 & 4.25 & 5.18 & \phantom{0}40 & 4.24 & 5.23 \\
\phantom{0}9 & 4.208 (0.100) & 5.072 (0.050) & 19 & 4.25 & 5.20 & \phantom{0}50 & 4.24 & 5.24 \\
10 & 4.221 (0.100) & 5.047 (0.050) & 20 & 4.25 & 5.20 & \phantom{0}60 & 4.24 & 5.25
\\
11 & 4.23 \phantom{0(0.100)}& 5.07 \phantom{0(0.100)} & 21 & 4.26 & 5.21 & \phantom{0}70 & 4.24 & 5.26 \\
12 & 4.24 \phantom{0(0.100)} & 5.10 \phantom{0(0.100)} & 22 & 4.26 & 5.21 & \phantom{0}80 & 4.24 & 5.26 \\
13 & 4.25 \phantom{0(0.100)} & 5.14 \phantom{0(0.100)} & 23 & 4.26 & 5.21 & \phantom{0}90 & 4.24 & 5.26 \\
14 & 4.25 \phantom{0(0.100)} & 5.16 \phantom{0(0.100)} & 24 & 4.26 & 5.22 & 100 & 4.24 & 5.26 \\
\hline
\end{tabular*}
\end{table}

\printaddresses

\end{document}